# Coherent beam combining of optical vortices

HOSSEIN FATHI,[1*] MIKKO NÄRHI[1], REGINA GUMENYUK[1,2]

[1]Laboratory of Photonics, Physics Unit, Faculty of Engineering and Natural Sciences, Tampere University, 33720 Tampere, Finland.
[2]Tampere Institute for Advanced Study, Tampere University, Kalevantie 4, Tampere 33100, Finland

*Corresponding author: hossein.fathi@tuni.fi

**We experimentally demonstrate the power scaling of optical vortices by the coherent beam combining, encompassing topological charges ranging from $\ell=1$ to $\ell=5$ realized on the basis of a Yb-doped fiber short-pulsed laser system. The combining efficiency varies from 83.2 to 96.9% depending on the topological charge and beam pattern quality generated by the spatial light modulators. These results open a pathway to high-intensity optical vortices with enormous potential applications in science and industry by utilizing advances in light-matter interactions.**

Optical vortices (OVs), spiral-shaped light beams with orbital angular momentum (OAM), have undergone substantial advancements and gained extensive interest since their introduction in 1989 [1-3]. By precisely tailoring the spatial structure of the beam, researchers can manipulate its properties and achieve functionalities beyond those of traditional Gaussian beams. OVs characterized by Hilbert factor $\exp(i\ell\varphi)$, $\ell$ is an integer number and $\varphi$ is the azimuthal angle, possess a spiral phase front and carry OAM equivalent to $\ell\hbar$ per photon, which is $\ell$ times larger than spin angular momentum (SAM), equivalent to $\pm\hbar$ per photon. These unique properties make them highly promising in a variety of applications such as optical tweezers [4-6], high-capacity optical communications [7-9], super-resolution microscopy [10, 11], and laser-plasma interaction [12, 13]. Moreover, improving the tunability of OVs, including spectral, temporal, topological charge (TC), chirality and singularity, spurred progress in various fields of advanced research [14-19].

There are two main ways to generate OVs [20], the first is by direct generation inside the laser cavity [21, 22], and the second by indirect mode conversion based on phase front modulators such as spatial light modulators (SLMs) as computer-generated holograms [23, 24], spiral phase plates [25], q-plates [26], and cylindrical lens [27].

There has been a sustained interest in generating high-power spatially structured beams to fully exploit their potential. For instance, in addressing the attenuation challenges encountered in free-space optical communication, enabling powerful optical trapping and manipulation, and facilitating high-power laser material processing. [4-13, 28, 29]. However, generating high-power OVs encounters limitations with all the previously mentioned conventional methods due to thermal damage of components. To address this issue, we employ coherent beam combining (CBC) as a versatile technique for power scaling OVs while preserving their desired characteristics for the first time, to the best of our knowledge. CBC is a method for increasing the power output of lasers by combining multiple laser amplifiers initially seeded by a common source [30]. The main idea involves splitting a seed laser into multiple replicas (N channels), amplifying each replica to its maximum power/energy through separate amplifier sections, and subsequently merging them into a single high-intensity beam while maintaining the beam's quality. Along with preserving the spatial properties, CBC also ensures that the spectral properties of the lasers are retained. This technique relies on establishing a phase relationship between the laser amplifiers, allowing them to operate effectively as a single laser amplifier. CBC technique has been fully established for both continuous wave (CW) and pulsed lasers, and to date, more than 100 kW average power, a few tens of mJ in pulse energy and a few tens of GW peak power have been experimentally demonstrated [31-33]. Furthermore, CBC reveals significant potential for scaling the number of channels, CBC of 107 beams has been experimentally demonstrated [34].

There are several reports on the generation of OVs using the CBC technique after the first demonstration in 2009 [35-40]. These studies consistently applied the CBC technique within a tiled-aperture configuration. This approach involves adjusting the intensity weights and the piston phase distributions of fundamental Gaussian array beams to create a helical phase structure, ultimately leading to generating the desired optical vortices in the far field. However, in this paper, we present a first-ever experimental demonstration of an active coherent beam combination of OVs via filled-aperture configuration. Our experiment involves the CBC of linearly polarized OVs with topological charges spanning from $\ell=1$ to $\ell=5$, all carried by short pulses. The beam combining efficiency depends on the quality and spatial pattern complexity of the generated vortices and can be as high as ~96%, roughly analogous to the combining efficiency for Gaussian beams.

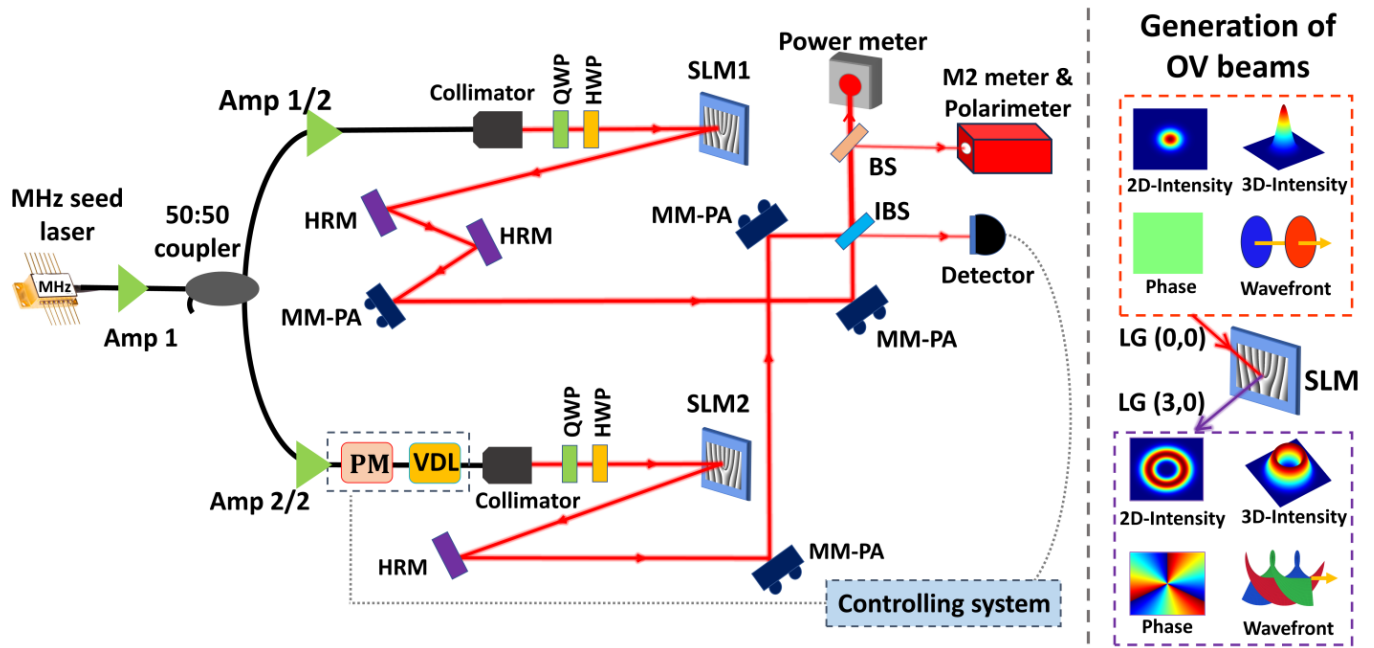

**Fig. 1.** Scheme of the experimental setup for the generation and coherent combination of optical vortices. The schematic process of the generation of the LG (3, 0) out of LG (0, 0) is shown on the right side of the picture. The Amp: Amplifier, PM: Phase Modulator, VDL: Variable Delay Line, QWP: Quarter Wave Plate, HWP: Half Wave Plate, SLM: Spatial Light Modulator, HRM: High Reflective Mirror, MM-PA: Motorized Mirror equipped with Piezo Actuators, IBS: Intensity Beam splitter, BS: Beam Sampler.

Fig. 1 illustrates the experimental setup employed for the generation and coherent combination of OVs. The scheme highlights the key components and techniques utilized in generating and combining OVs. The seed source consists of an ultrafast gain-switched laser that emits pulses at a repetition rate of 20 MHz with a pulse duration of 55 ps (±5 ps). Following amplification in the initial stage, the seed laser is split into two channels using a 50:50 fiber coupler. Subsequently, each channel undergoes a second amplification stage, delivering an average power of up to 100 mW. The second channel is equipped with a $LiNbO_3$ electrooptic phase modulator (PM) to precisely control and adjust the phase of the OVs, ensuring a stable and coherent combination. To achieve an efficient coherent combination, a variable delay line (VDL) is incorporated to equalize the optical path lengths of the OVs. Using two collimators at the end of each channel, the output beams were collimated to propagate in free space. Two Spatial light modulators (SLMs) are utilized to introduce the desired phase patterns, thereby generating Laguerre-Gaussian (LG) optical vortex beams with different TCs ($\ell$= 1-5 and $p$=0). For illustrative purposes, the schematic process for generating LG (3, 0) out of LG (0, 0) is depicted on the right side of Fig. 1. Since the SLMs are polarization-dependent, we utilize a pair of quarter-wave plate (QWP) and half-wave plate (HWP) to achieve the optimal polarization state. Each generated OV, with an output power of 50 mW, is then directed towards a set of high reflection mirror (HRM) mirrors that steer the beams into the combining elements. To ensure precise spatial overlapping and maximize the combining efficiency, two high-reflection motorized mirrors, equipped with piezo actuators were utilized in each channel. These actuators enable fine adjustments of the mirror positions, ensuring accurate alignment of the OVs. The filled-aperture configuration, utilizing near-field beam combining, has been employed in this setup. The combining element is an intensity beam splitter (IBS), which allows the two OVs to spatially overlap and combine coherently. The coherently combined beam emerges from a predetermined port of the intensity beam splitter. Using a beam sampler (BS), a portion of the combined beam is extracted for measuring the beam quality factor. The uncombined light that emerges from the idler port of the IBS is utilized as feedback for the active control system. The beam phase control system employed here is a customized commercial feedback loop system, based on a top-of-fringe stabilization technique (Laselock, TEM Messetechnik).

Through an extensive experimental investigation, five distinct linearly polarized Laguerre-Gaussian optical vortices, denoted as LG ($\ell$=1-5, $p$=0) and Gaussian beam ($\ell$=0, $p$=0) were coherently combined. The LG (0, 0) beams were coherently combined with an efficiency of 98.3% for the reference of the technique. The first five Laguerre-Gaussian OVs, each possessing distinct topological charges of $\ell$ = 1, 2, 3, 4, and 5, were combined with corresponding efficiencies of 95.5%, 86.9%, 83.2%, 78%, and 68.1%. Combining efficiency is calculated by dividing the power of the combined beam by the sum of the output power of each laser beam. The decrease in combining efficiency for high-order OVs can predominantly be attributed to the imperfection of the generated OVs by SLMs, as the field experiences some undesirable distortions during SLM-induced modulation. Furthermore, differences in collimation quality resulting in different divergences for the two beams, as well as an imbalance in the free-space optical paths, could also lead to a decrease in efficiency.

Fig. 2 illustrates the beam profiles of five distinct linearly polarized Laguerre-Gaussian optical vortices and a Gaussian beam for two different channels and their coherent combination output, along with the corresponding phase patterns applied on SLM to generate them from the LG (0,0). They show the time-varying beam profiles of the combined beams when the phase control system is on (coherently combination) and off (random combination). Additionally, the videos include separate beam profiles for each individual channel beam of the replica. Other LG OVs ($\ell$ =2-4, $p$ =0) demonstrated similar behaviour during combination. The output beam qualities of all combined LG OVs were analyzed. We would like to note that the combined beam demonstrated higher uniformity of intensity distribution. This could be interpreted as an improvement of the beam quality by CBC. Fig. 3 depicts the ISO 11146-compliant M² measurement of the combined beams utilizing the 4$\sigma$-method, along with the corresponding combining efficiencies. Fig. 3(b), 3(c), and 3(d) illustrate the M² measurement of the coherently combined beams for LG (0,0), LG (1,0), and LG (5,0), respectively, accompanied by the corresponding near-field beam profiles as insets. We present three of six beams as examples, and other OVs demonstrated similar performance. The typical optical spectrum of both channels and the combined one for LG (1,0) are depicted in Fig. 4(a). For further analysis, we characterized the optical pulse durations of both individual channels and the combined output in the time domain using a 25-GHz photodetector, as illustrated in Fig. 4(b). The full width at half maximum (FWHM) pulse duration was 55±5 ps. We have not noticed any spectral or temporal profile changes for other LG modes. Polarization and output power stability assessment was conducted on the combined OV beam of LG (1,0) over 20 minutes using a commercial polarimeter (PAX1000IR2/M), presenting a common performance for all OV beams. The results corresponding to this assessment are shown in Fig. 5, providing a clear illustration of the excellent polarization and power stability of the CBC system. The power fluctuation was less than 3.7%, while DOP variations did not exceed 0.2%. The ellipticity of the combined beam demonstrated negligible changes.

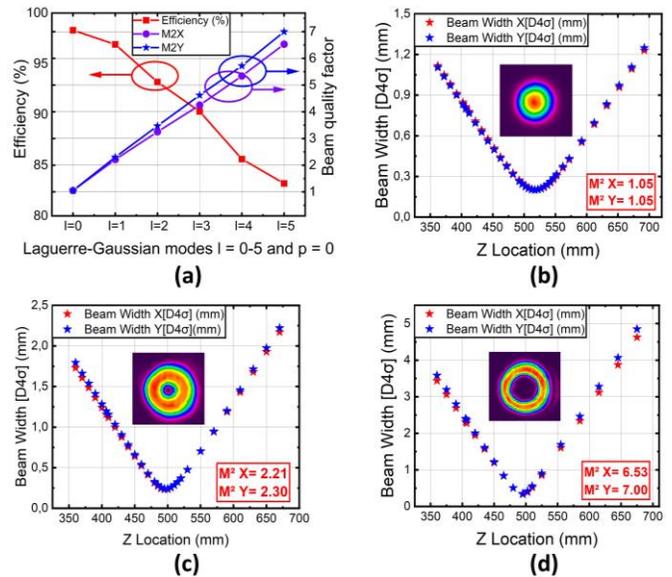

**Fig. 3.** Analysis of beam quality and combining efficiency of the combined beams. (a): ISO 11146-compliant M2-measurement of the combined beams with the 4$\sigma$-method with corresponding combining efficiencies, (b), (c), and (d): M2-measurement of the combined beams of LG (0.0), LG (1.0), and LG (5.0), respectively (Insets: near field beam profiles of the combined beams).

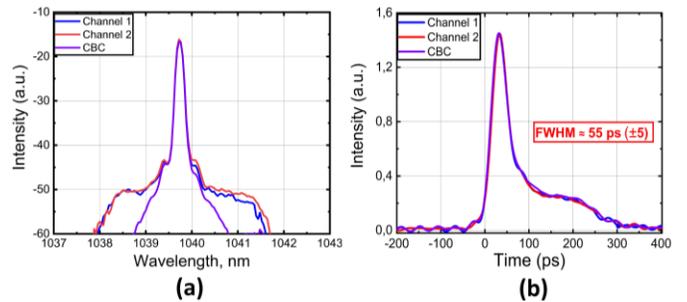

**Fig. 4.** Optical spectrum and pulse duration measurements of the output of channel 1, channel 2 and the combined output of both channels for LG (1,0). (a) Normalized optical spectra, (b) time-domain envelope measured using a 25-GHz photodetector.

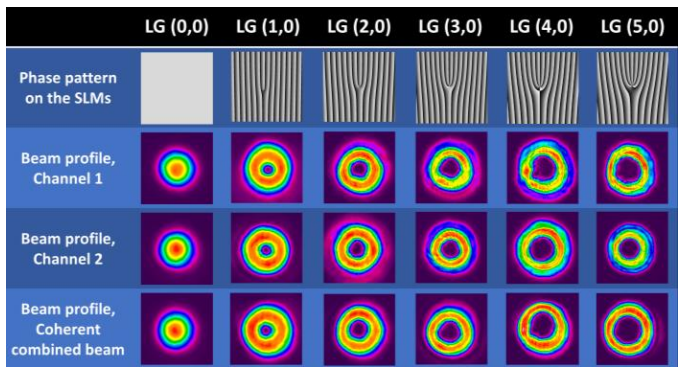

**Fig. 2.** Profiles of five distinct linearly polarized Laguerre-Gaussian optical vortices, LG OVs ($\ell$=1-5, p=0) and a Gaussian beam ($\ell$=0, p=0), for two different channels and their coherent combination, along with the corresponding phase patterns applied on SLM.

In conclusion, we presented the first experimental demonstration, to the best of our knowledge, of coherent beam combining for Laguerre-Gaussian optical vortices, achieving high combining efficiencies of up to ~96% in a filled-aperture configuration. Remarkably, the combined beam demonstrated improved uniformity of the spatial intensity distribution for all OVs, which could be interpreted as improved beam quality. These results mark a pioneering achievement in the potential for power/energy scaling of optical vortices in short-pulsed laser systems, thereby paving the way for exploring novel applications in the field of high-intensity light-matter interactions. The CBC technique is highly scalable, revealing promising benefits for fields that require high-power OAM beams, as a case in point, overcoming signal attenuation in long-distance free-space optical communications. To maximize combining efficiency, it is essential to generate high-quality OVs, especially for higher-order OVs, where efficiency decreases. Therefore, attention should be given to the quality of SLMs or phase plates used, as well as free-space beam propagation principles. Ongoing research and development endeavours in this domain will further refine the implementation of CBC for extremely high-power OVs, unlocking new possibilities for various scientific and technological applications.

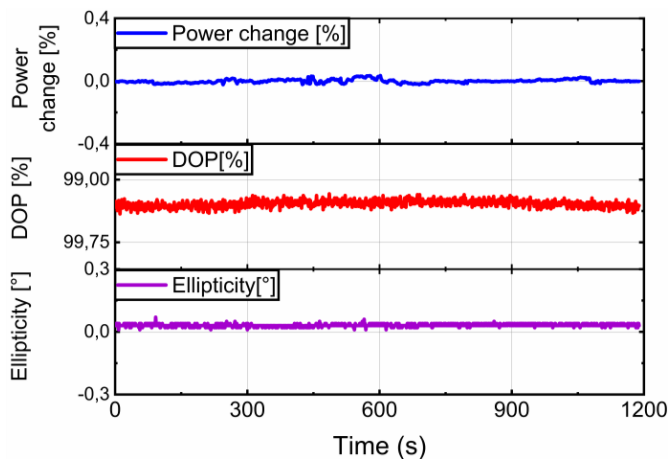

**Fig. 5.** Polarization and output power measurements stability of the output of coherent beam combination of two LG (1, 0) beams. (a) Output power stability, (b) degree of polarization (DOP), and (c) the Ellipticity.

**Funding.** Horizon Europe EIC Pathfinder-OPEN (101096317), Research Council of Finland (320165).

**Acknowledgements.** The authors thank Rafael Barros, Hassan Asgharzadeh, and Ebrahim Aghayari for fruitful discussions.

**Disclosures.** The authors declare no conflicts of interest.

**Data availability.** Data underlying the results presented in this paper are not publicly available at this time but may be obtained from the authors upon reasonable request.